# Resolving diverse oxygen transport pathways across Sr-doped lanthanum ferrite and metal-perovskite heterostructures

S.D. Taylor,[1,+,*] K.H. Yano,[2,+] M. Sassi[1], B.E. Matthews,[2] E.J. Kautz,[2] S.V. Lambeets,[1] S. Neumann,[2] D.K. Schreiber,[2] L. Wang,[1] Y. Du,[1] S.R. Spurgeon[2,3,*]

1. Physical and Computational Sciences Directorate, Pacific Northwest National Laboratory, Richland, WA 99352 USA
2. Energy and Environment Directorate, Pacific Northwest National Laboratory, Richland, WA 99352 USA
3. Department of Physics, University of Washington, Seattle, WA 98195 USA

[+]First coauthors.

*Corresponding authors: sandra.taylor@pnnl.gov; steven.spurgeon@pnnl.gov

# Abstract

Perovskite structured transition metal oxides are important technological materials for catalysis and solid oxide fuel cell applications. Their functionality often depends on oxygen diffusivity and mobility through complex oxide heterostructures, which can be significantly impacted by structural and chemical modifications, such as doping. Further, when utilized within electrochemical cells, interfacial reactions with other components (e.g., Ni- and Cr-based alloy electrodes and interconnects) can influence the perovskite's reactivity and ion transport, leading to complex dependencies that are difficult to control in real-world environments. Here we use isotopic tracers and atom probe tomography to directly visualize oxygen diffusion and transport pathways across perovskite and metal-perovskite heterostructures, i.e., (Ni-Cr coated) Sr-doped lanthanum ferrite (LSFO). Annealing in $^{18}O_{2(g)}$ results in elemental and isotopic redistributions through oxygen exchange (OE) in the LSFO while Ni-Cr undergoes oxidation via multiple mechanisms and transport pathways. Complementary density functional theory (DFT) calculations at experimental conditions provide rationale for OE reaction mechanisms and reveal a complex interplay of different thermodynamic and kinetic drivers. Our results shed light on the fundamental coupling of defects and oxygen transport in an important class of catalytic materials.

# Introduction

Perovskite structured transition metal oxides (general chemical formula of $ABO_3$) are widely studied in chemistry and condensed matter physics, owing to their strong coupling among lattice, spin, and orbital degrees of freedom.[1] These crystals can accommodate a variety of property-defining cation species, giving rise to diverse electronic, magnetic, and optical behavior.[2] For instance, their catalytic activity and properties can be significantly influenced by substitution or partial substitution of the A- and/or B-site cations.[3–6] Among the many perovskites being pursued for catalytic applications, Sr-doped lanthanum ferrites ($La_{1-x}Sr_xFeO_3$; LSFO) have attracted particular attention for photocatalytic water splitting,[7–10] with Fe as the B-site transition metal cation driving selective oxidation. The $La^{3+}$ cations are substituted by cations in a lower oxidation state (i.e., $Sr^{2+}$), leading to the partial oxidation of the B cations to a higher oxidation state and/or to the formation of oxygen vacancies, which results in better catalytic activity.[10] The perovskite's ability to accommodate a range of substituents and dopants provides significant flexibility in its composition and associated oxidation state. This tunability in turn enables tailoring of the perovskite's physicochemical properties for various applications such as cathode materials in solid oxide fuel cells (SOFCs), catalysts and oxygen carriers in heterogeneous catalysis, oxygen separation membranes, and solid-state gas sensors.[11]

The mobility of oxygen ions is a key parameter affecting a perovskite's reactivity and functionality across these applications.[12] Diffusion through the oxide bulk follows a vacancy-mediated mechanism, where transport occurs by discrete hops of oxygen anions to neighboring vacancies.[10] Diffusion rates can be further manipulated by modifying vacancy or other defect populations via cation doping.[13–15] Further, as mixed ionic-electronic conductors, oxygen from the environment can reversibly adsorb/desorb into the lattice and exchange by continuous changes in oxidation state, without changes to the bulk crystal structure.[10,16,17] However, the oxygen exchange (OE) reaction itself consists of numerous elementary steps and reaction sequences, such as oxygen adsorption, reduction and dissociation, diffusion of the disassociated species, and subsequent incorporation into the host (cathode/electrolyte) lattice,[18] which have yet to be fully understood.

Within designed electrochemical cells, ion transport is also influenced by synergistic reactions with the various components of the electrode and interconnects. For instance, in the design of intermediate temperature (600-800°C) SOFCs, chromia-forming stainless steel interconnects are desirable due to the higher electronic and thermal conductivity, lower cost, and easier fabrication than traditional ceramic parts.[19–21] However, a significant challenge in their application is the severe degradation of the cathode performance resulting from poisoning via Cr-species evaporated from the interconnect alloy oxide scale. Deposition of Cr-species blocks active sites on the electrode surface, negatively affecting charge transfer and oxygen diffusion,[19] although the specific controlling mechanisms are convoluted and poorly understood.[22] Knowledge of oxygen reactivity and transport pathways in the oxide catalyst and relative to components of the electrochemical cell can provide fundamental insight into degradation mechanisms, guiding cell design to optimize performance.

In this study, we directly resolve oxygen transport pathways across (metal-)perovskite heterostructures at the nanoscale for insight into diffusivity and surface exchange reactions. To do so, oxygen diffusion was studied within a model LSFO system ($La_{0.5}Sr_{0.5}FeO_3$ grown on (001) $[LaAlO_3]_{0.3}[Sr_2AlTaO_6]_{0.7}$, [LSAT]) and across a heterogeneous metal-LSFO system (Ni- and Cr-metal). We employed recently developed isotopic tracer techniques with atom probe tomography (APT), a powerful technique used to resolve elemental and isotopic distributions in three dimensions (3D) with sub-nanometer resolution.[23–26] LSFO serves as a model system to study vacancy-mediated oxygen transport in the bulk, as rapid diffusion was previously measured although this has not been directly resolved at the nanoscale.[13,27] Thin films of Ni and Cr were also deposited on top of the LSFO as surrogate interfaces for metallic anodes and interconnects in electrochemical cells.[28] APT specimens were prepared and annealed in gaseous $^{18}O_2$ using the *in situ* atom probe (ISAP) method.[29] After annealing, local changes in the elemental and isotopic distributions were readily observed by APT, informing unique transport pathways across the stack. These observations are key to understanding mass transport phenomena controlling the degradation, rejuvenation, and stability of complex perovskite catalysts and oxide heterostructures.

## Methods

### Materials synthesis

$La_{1-x}Sr_xFeO_3$ (x = 0.5) films nominally ~9 nm thick (~22 unit cells) were epitaxially grown on (001)-oriented LSAT substrates using pulsed laser deposition; the growth details have been described elsewhere.[9] In brief, the laser pulse (248 nm) energy density was ~2 J/cm² and the repetition rate is 1 Hz. During the deposition, the substrate was kept at 700 °C under an oxygen partial pressure of 10 mTorr. After deposition, the samples were cooled to room temperature in 10 mTorr oxygen. Cr- and Ni-layers (~15 nm and ~30 nm, respectively) were then deposited on the LSFO surface via ion beam sputtering deposition (IBSD) at room temperature.

### Annealing experiments

To probe oxygen transport across this system, an isotopic tracer was used during material processing to couple with APT analysis. APT specimens were annealed in $^{18}O$-enriched oxygen gas (400°C, 4h, 3 mbar $^{18}O_2$; 99 atomic % (at.%) purity, Sigma Aldrich) using the in ISAP method, detailed elsewhere.[29] This method allows for the direct observation of sub-nanometer scale element redistributions at the specimen surface.[30–32] Briefly, this approach involves the direct exposure of APT specimens to a controlled gaseous environment at select pressures, temperatures, and times in a chemical reactor chamber attached directly to a commercial local electrode atom probe (LEAP) system. Following annealing the specimens were transferred directly into the APT analysis chamber under ultrahigh vacuum and analyzed. The conditions were chosen such that oxygen transport across the thin film (~10 nm) could be resolved by APT, based on extrapolations from previous studies; i.e., oxygen diffusion at 400°C is estimated to be ~$10^{-18}$ m² s⁻¹ or 1 nm² s⁻¹ when extrapolating diffusion measurements in $La_{0.6}Sr_{0.4}FeO_3$ from 900 – 1100°C.[13]

## APT analyses

Samples were prepared for scanning transmission electron microscopy (STEM) and APT using conventional focused-ion beam scanning electron microscopy techniques (FIB-SEM; FEI Quanta 3D-FEG or Helios NanoLab dual-beam microscopes).[33,34] In particular, the APT specimens were prepared such that the final tip geometry consisted of Ni at the tip apex, underlain by Cr, LSFO, and LSAT, in sequence (see APT specimen preparation protocol in Supplementary Information (SI); Figure S1). Microstructural and compositional characterizations of the thin film by both STEM and APT were done to get baseline measurements of the as-grown specimens.

APT samples were analyzed in a CAMECA LEAP 4000XHR at a base temperature of 40 K in laser-assisted field evaporation ($\lambda = 355$ nm) at pulse rates of 125 and 200 kHz. The laser pulse energy was set between 80 and 150 pJ and the detection rate was maintained at 0.002 detected ions per pulse by varying the applied volage. The 3D APT reconstructions were done using the Integrated Visualization and Analysis Software (IVAS 3.8.5a45) with the shank angle approximation. APT reconstructions were scaled using STEM measurements of the film thicknesses and the interplanar spacing of the LSFO atomic bilayers that were also resolvable using spatial distribution maps (SDMs). The oxygen bilayer spacing of LSFO as calculated by density functional theory (DFT) is, on average, 1.945Å in the [001] direction, and APT field evaporation and thus reconstruction captures every other of these bilayers (3.89Å). Using this scaling the films' thicknesses were consistent with that measured by STEM. Elemental and isotopic analyses were achieved through the careful assignment of ionic species best representative of the different phases (see SI for more detail).

Two APT specimens were analyzed for each condition (i.e., as-grown, annealed perovskite, and annealed metal-perovskite). Elemental concentrations were used to monitor potential phase transformations in the specimens after annealing in $O_2$. Isotopically-resolved measurements were used to follow the provenance of oxygen from the reactor chamber ($^{18}O$) across the specimen as the as-grown oxides were $^{16}O$-dominant (natural abundance (NA) for O = 99.8% $^{16}O$, 0.2% $^{18}O$) and interpret oxygen ingress and mobility from the reactor environment. Elemental and isotopic compositions were shown to be consistent between duplicate APT specimens in the same condition (Table S1). One specimen from each experiment is shown in the main text as the representative sample; elemental and isotopic analyses of the remaining specimens are provided in the SI (Figure S2).

## STEM analyses

STEM high-angle annular dark field (STEM-HAADF) images were acquired on a probe-corrected JEOL GrandARM-300F microscope operating at 300 kV, with a convergence semi-angle of 29.7 mrad and a collection angle range of 75 – 515 mrad.

## Theory calculations

The VASP package[35] was used to perform density functional theory (DFT) simulations of defective and Sr-doped LSFO materials. All the simulations used the generalized gradient approximation (GGA) as parametrized in the PBEsol functional[36] in combination with the Hubbard correction[37] ($U_{eff} = U - J = 4$ eV) to better describe the Coulomb repulsion of the $3d$ electrons of the Fe atoms.[38] All the simulations used a cutoff energy of 550 eV and a 2×2×2 Monkhorst-Pack $k$-points mesh to sample the Brillouin zone. The total energy was converged to $10^{-6}$ eV/cell, the force components were relaxed to $10^{-5}$ eV/Å, and all the simulations used spin-polarization.

The generation of a 50% Sr-doped LSFO (i.e., $La_{0.5}Sr_{0.5}FeO_3$) unit cell used as a starting structure a 2×2×2 supercell (160 atoms) of pure $LaFeO_3$ (LFO) compound. The relaxation of the orthorhombic (space group *Pbnm* #62) LFO as a G-type anti-ferromagnetic materials yielded lattice parameters $a = 5.537$ Å (-0.72%), $b = 5.537$ Å (-0.60%), and $c = 7.793$ Å (-0.72%), in good agreement with the experimental lattice parameters $a = 5.55$ Å, $b = 5.57$ Å, and $c = 7.85$ Å.[39] The structure of a 50% La/Sr mixed compound was generated using the Special Quasirandom Structure (SQS) code available from the Alloy Theoretic Automated Toolkit (ATAT).[40] Once the lattice parameters and coordinates of the 50% Sr-doped structure were optimized, one oxygen vacancy was introduced into the structure and only the coordinates were relaxed while the lattice parameters were kept fixed at their optimized values.

The temperature and $O_2$ partial pressure ($p_{O_2}$) dependence of the Gibbs free energy, $\Delta G_f(T, p_{O_2})$, of oxygen vacancy was determined by *ab initio* thermodynamics simulations, following the equation:

$$\Delta G_f(T, p_{O_2}) = \left(E_{V_O}^T + E_{V_O}^{ZPE} + \Delta\mu(T)_{V_O}\right) - \left(E_{perf}^T + E_{perf}^{ZPE} + \Delta\mu(T)_{perf}\right) + \frac{1}{2}(E_{O_2}^T + E_{O_2}^{ZPE} + \Delta\mu_{O_2}(T, p_{O_2})) \quad (1)$$

where $E_{V_O}^T$ and $E_{perf}^T$ are the DFT total energies of the $La_{0.5}Sr_{0.5}FeO_3$ solid systems with and without oxygen vacancy, $E_{V_O}^{ZPE}$ and $E_{perf}^{ZPE}$ are the zero-point-energy of the defective and defect-free system. $E_{O_2}^T$, $E_{O_2}^{ZPE}$, and $\Delta\mu_{O_2}(T, p_{O_2})$ are respectively the total DFT energy, the zero-point-energy, and the temperature and $O_2$ partial pressure dependent chemical potential of oxygen. The DFT total energy of molecular $O_2$ was corrected to the experimental atomization energy of the gaseous species[41] leading to an energy correction of 1.482 eV. In order to account for temperature effect in $La_{0.5}Sr_{0.5}FeO_3$, $\Delta\mu(T)_{V_O}$ and $\Delta\mu(T)_{perf}$ are the temperature-dependent chemical potentials of the system with and without an oxygen vacancy. All the temperature-dependent chemical potentials were calculated using the following relation:

$$\Delta\mu(T) = (H(T) - H°(298.15)) - TS \quad (2)$$

where $H(T)$ and $H°(298.15)$ are the system enthalpy at a temperature $T$ and at $T=298.15$ K, and $S$ is the entropy. In the simulations, the phonon frequencies were calculated using the Phonopy code.[42]

The simulations of O vacancy migration pathways used the climbing image nudged elastic band method (CINEB) as implemented in the Transition State Tools for VASP (VTST).[43]

# Results and Discussion

## Baseline characterization of as-grown sample

The baseline microstructure of the heterostructure was established by high-resolution STEM-HAADF imaging, as shown in Figure 1. Atomic resolution imaging confirms the epitaxial and single crystal nature of the LSFO thin films on the highly crystalline-LSAT substrate. The sputtered Ni and Cr metal layers are nanocrystalline, as expected. The LSFO film structure is largely uniform with few to no microscopic defects (with the exception of some steps near the LSFO-LSAT interface) (Figure 1b), though oxygen vacancy concentrations are expected to be significant at this level of Sr-doping.[14]

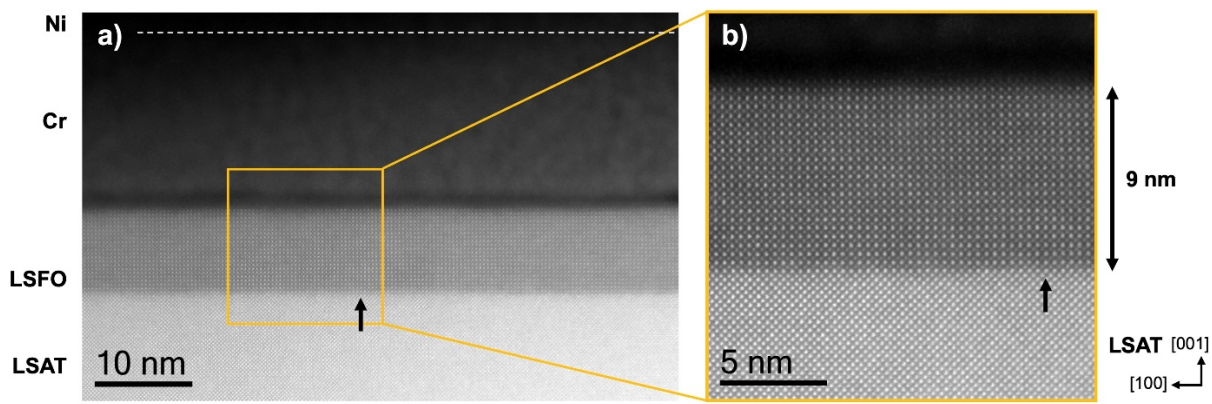

*Figure 1: Cross-sectional STEM-HAADF imaging of a) the Ni-Cr-LSFO-LSAT heterostructure taken along the LSAT [010] zone, and b) the LSFO and LSAT structure with black arrows indicating steps in the substrate.*

The elemental and isotopic composition of the metal-perovskite stack was characterized by APT (see SI for more details on elemental analyses, Figure S2 and Table S1). 3D chemical reconstructions clearly show the unique stack geometry and chemistry of the metal-perovskite system (Figure 2a). The baseline compositions of the phases in the as-grown specimens are in reasonable agreement with that expected (Figure 2b, Table S1,). That is, the Ni and Cr coatings are largely metallic (77.7 ± 2.5 at.% Ni and 70.6 ± 1.5 at.% Cr, respectively, based on the average composition and standard deviation across the two APT tips measured). Oxygen is present in both layers (13.3 ± 1.5 at.% and 21.6 ± 1.2 at.% in both Ni and Cr layers, respectively), which is present in the IBSD chamber during deposition.[23] Aluminum, an impurity introduced by Al components in the sputter system, is also present at low concentrations (6.8 ± 1.5 at.% and 6.6 ± 0.25 at.% in Ni and Cr layers, respectively).

Quantification of the as-grown LSFO composition shows that Fe is the primary cation (27.2 ± 0.1 at.% Fe) followed by Sr and then La (18.5 ± 0.4 at.% and 10.7 ± 0.5 at.%, respectively). The measured O concentration is only 43.3 ± 0.9 at.% versus the expected concentration of 60 at.% O of the $ABO_3$ stoichiometry. While some of this O deficiency could reflect oxygen vacancies due to the Sr-doping, it is important to note that oxygen quantification is inherently challenging in APT[44,45], especially in Fe-base oxides.[23,44,46–48] We can nonetheless follow changes in concentrations before and after annealing to determine relative composition changes. The APT-measured as-grown LSAT composition (7.6 at.% La, 14.1 at.% Sr, 10.0 at.% Al, 8.5 at.% Ta, and 58.7 at.% O, based on the composition from one APT specimen) is in reasonable agreement with expectations (3.6 at.% La, 16.4 at.% Sr, 11.8 at.% Al, 8.2 at.% Ta, and 60 at.% O). APT analyses along the defined [001] direction of the LSFO film show a homogeneous composition along the depth of each phase, with little to no elemental intermixing as demarcated by the sharp interfaces between phases (Figure 2a-b).

The oxygen isotopic composition of each phase and across the heterostructure was determined using a single, consistent species, i.e. $O^{1+}$ species (16 – 18 Da). This avoided potential interferences given the multi-isotopic and elemental nature of this system (see Methods and SI for details on $^{18}O$ analytical protocols), The oxygen composition is expressed as the fraction of $^{18}O$ relative to $^{16}O$ and $^{18}O$ ($f^{18}O$).[23] The average $f^{18}O$ measurements for the bulk Ni, Cr, LSFO, and LSAT was 0.05 ± 0.02, 0.04 ± 0.00(1), 0.01 ± 0.00(4), and 0.01 respectively (Figure 2c, Table S1). These $f^{18}O$ measurements within each phase of the as-grown specimen are within reason of that expected, i.e., near NA (0.002); the presence of hydrogen in the APT analysis chamber leads to the slight deviations.[49] Nonetheless, these measurements show the oxygen isotopic composition within the as-grown specimen is effectively at NA. As shown shortly, following annealing, changes in $f^{18}O$ are significant and thereby can be used to reliably infer oxygen diffusivity and reactions across the stack.

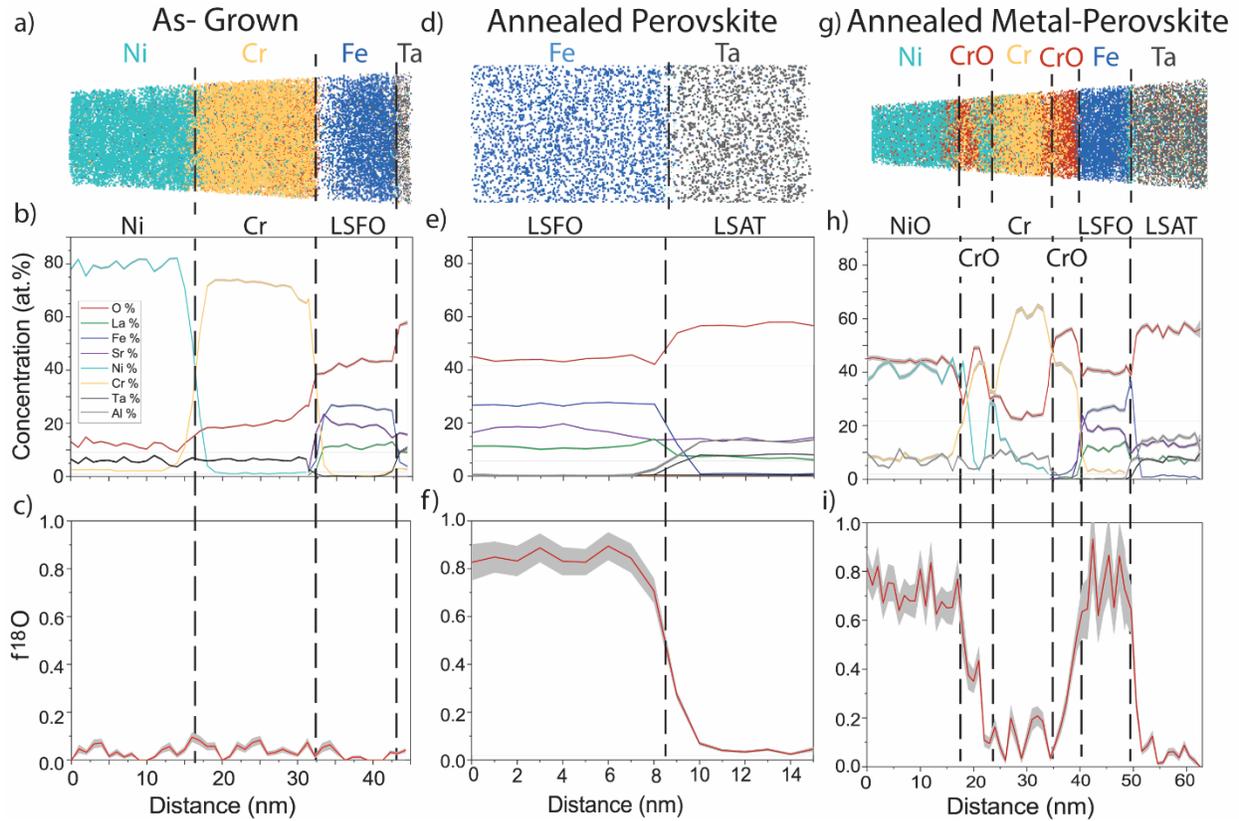

Figure 2: APT reconstructions, elemental profiles, and $f^{18}O$ for the (a-c) as-grown vs. (d-f) annealed perovskite vs. (g-i) annealed metal-perovskite specimens. Shading is used to highlight the uncertainty associated with the counting statistics (1σ, see Methods).

To compare the behavior and mobility of oxygen in isolated perovskites to those in the presence of the metallic overlayers, APT was used to further prepare specimens prior to reaction. For some APT specimens, the Ni-Cr overlayers were field evaporated, leaving tips with only LSFO-LSAT remaining. The remaining specimens, APT specimens with Ni-Cr-LSFO-LSAT configuration, remained intact. These two specimen geometries are referred to as annealed perovskite and annealed metal-perovskite, respectively. The annealed LSFO-LSAT specimen enables us to isolate and infer oxygen mobility based on properties inherent to perovskites, which are further guided by complementary *ab initio* simulations performed here and observations from previous studies. In turn, these analyses also provide insight into unique oxygen transport mechanisms for the annealed metal-perovskite specimens.

## Oxygen mobility in the annealed-perovskite system

Following annealing of the perovskite tips in $^{18}O_2$, the elemental and isotopic compositions were assessed to determine whether the phase was altered. The elemental compositions of the annealed LSFO and LSAT layers are consistent with their as-grown counterparts (Figure 2d-f, Table S1). Notably, the O concentration in the annealed LSFO is within variability of the as-grown measurements (43.8 ± 0.3 at.% vs. 43.3 ± 0.9 at.%, respectively), indicating there is no net addition or removal of oxygen during reaction.

Though elemental analyses indicate phase alteration does not occur, the oxygen isotopic composition demonstrates that lattice natural isotopic abundance oxygen in the LSFO nearly completely exchanges with the $^{18}O$-enriched environmental $O_2$ (Figure 2f, Table S1). Quantitatively, the LSFO $f^{18}O$ increases from 0.01 ± 0.0(04) in the as-grown specimen to 0.84 ± 0.01 after annealing. In comparison LSAT does not experience OE, remaining near NA ($f^{18}O$ = 0.04). Additionally, isotopic measurements along [001] indicate OE is homogeneous along the depth of the LSFO film. Elemental composition analyses also show that the LSFO and LSAT exhibit no evidence of phase transformations.

While OE in perovskites has not been measured previously at these modest temperatures, the extent of exchange seems reasonable based on rough predictions for oxygen diffusivity in LSFO and surface exchange to occur. Arrhenius extrapolations of $La_{0.6}Sr_{0.4}FeO_3$ estimate oxygen diffusivity at 400°C to be ~$10^{-18}$ m² s⁻¹ or 1 nm² s⁻¹.[13] Using that estimate, the characteristic diffusion distance for the 4h anneal would be around 240 nm, significantly longer than our film thickness of 10 nm, consistent with the total OE we observe. Additionally, the needle-shaped APT specimen geometry and large surface area exposed could facilitate OE via surface exchange. For example, LFO annealed in $^{18}O_{2\,(g)}$ at > 900°C showed the most OE occurs near the surface and decreases into the bulk, based on limited observations from lower-resolution isotopic exchange depth profile measurements with secondary ion mass spectrometry (i.e., > 100 nm).[27]

## Oxygen exchange mechanisms

Oxygen transport and exchange in LSFO is ultimately controlled by the behavior and concentration of oxygen vacancies.[18] Our experimental observations are consistent with this (e.g., the highly-crystalline LSAT is unreactive while extensive oxygen diffusion and exchange in LSFO reflects inherently high defect concentrations). To support the experimental observations, atomistic modeling was used to further probe the vacancy-mediated transport mechanisms and driving forces at play. Specifically, the thermodynamic stability and migration energy barriers of oxygen vacancies in LSFO were simulated and quantified using DFT. The thermodynamic stability was first probed using *ab initio* thermodynamics calculations that can incorporate the effects of various experimental conditions (e.g., oxygen partial pressure, temperature, chemical potential differences between oxygen isotopes, and Sr concentrations). Subsequently the energetic drivers and migration barriers for vacancy-mediated transport, thought to influence the incorporation/removal of lattice oxygen and diffusion, were further evaluated.

The calculation of the Gibbs free energy of formation of an oxygen vacancy in LSFO indicates that heating increases the instability of vacancies, which may be a potential driver for oxygen incorporation into the lattice (Figure 3a). The degree of instability introduced by heating is also affected by the oxygen partial pressure, such that an increase in $pO_2$ further increases the instability of O vacancies. At the oxygen partial pressure used in this study ($pO_2$ = 3 mbar), O vacancies are found to be 1.31 eV less stable at 670 K than at ambient temperature. This thermodynamically favors oxygen ingress from the environment to occupy O vacancies. Countering this tendency, the vacancy concentration in the material is also restrained and maintained by the perovskite composition (i.e., Sr doping); our experimental observations are nominally consistent with this notion, as the oxygen concentration (an indirect indication of vacancy content) does not change during reaction. Thus, while annealing makes O vacancies less energetically favorable thermodynamically, a steady-state vacancy concentration is maintained throughout the reaction.

To understand oxygen diffusion via vacancy-hopping mechanisms, the energetics for O vacancy migration in the perovskite lattice were calculated using the climbing image nudged elastic band (CINEB) method. To quantify the impact of Sr doping on the O vacancy migration energy barrier as a function of the Sr/O-vacancy separation distance (Figure 3b), we used a supercell where one La atom was substituted by a Sr atom and compared it to O vacancy migration in pure LFO. In the case of pure LFO, the highest calculated energy barrier is 0.73 eV, in good agreement with previously reported experimental and theoretical values of ~0.70 – 0.80 eV.[12,13,50] However, when Sr is present in the lattice, the migration of a 1$^{st}$ neighbor O vacancy away from Sr must overcome a first energy barrier of 0.80 eV, compared to 0.62 eV for pure LFO. While subsequent energy barriers along the pathway are smaller, ranging from 0.56 eV to 0.69 eV, this indicates that O vacancies are more strongly associated with Sr. The trend obtained is in good agreement with the literature, suggesting that the energy barrier for the migration of the O vacancy increases with the Sr fraction.[50–52] However, these high energetic barriers seemingly contradict observations from diffusion studies, as oxygen diffusion increases with Sr doping (e.g., bulk diffusion in $La_{1-x}Sr_xFeO_3$ [x = 0.4, 1000°C] was measured to be 4 – 5 orders of magnitude faster than that in pure LFO).[13,50–52] This suggests that the impact of Sr doping on the concentration of O vacancies is a more important factor driving oxygen migration than its associated cost to the energy barrier. Therefore, high vacancy concentrations predominantly control oxygen diffusivity in LSFO, irrespective of the energetic barriers for vacancy migration, and enable long-range oxygen diffusion and migration.

In order to provide a rationale for the significant OE observed, bulk thermodynamic calculations for LSFO composed of either $^{16}O$ and $^{18}O$ were performed as function of temperature to evaluate the impact of isotope substitution on the chemical potential of LSFO. At 400°C, a difference of 20 meV/formula unit (1 formula unit = $La_{0.5}Sr_{0.5}FeO_3$ or 6.67 meV/O atom) in favor of the $^{18}O$-LSFO compound is obtained (Figure 3c). Based on this difference, the calculation of the Boltzmann factor to approximate the fraction of $^{18}O$ yields to 70% if we consider the energy gain per formula unit or 89% if we use the energy gain per O atom. On average, this approximation suggests that LSFO would favor the near complete replacement (~80%) of lattice $^{16}O$ by $^{18}O$, which is consistent with our experimental observations. The influence of isotopic mass in combination with the thermodynamic instability of vacancies at 400°C and $pO_2$ = 3 mbar highlights some of the potential driving forces for OE.

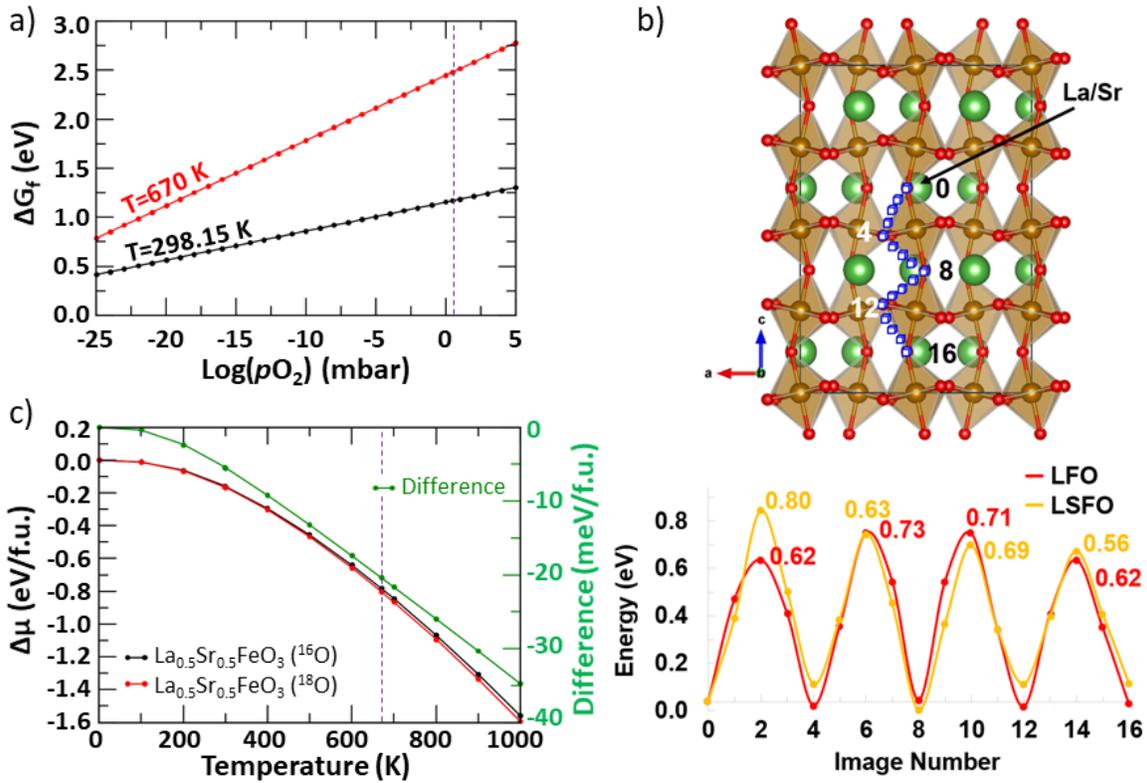

Figure 3: a) Gibbs free energy of O vacancy formation in LSFO (x=0.5) as function of $pO_2$ for temperatures of 25°C (298K) and 400°C (670K). The vertical dashed line indicates the experimental $pO_2$ (3 mbar). b) Impact of a single Sr substitution on the migration energy barrier of O vacancy. The O vacancy pathway investigated is indicated by blue squares. When Sr is introduced, the initial image configuration (position 0) corresponds to a O vacancy being 1st neighbor with the Sr species. The energy barriers have been calculated relative to the local minima just prior to the transition state. c) Comparison of the chemical potential (per formula unit) of LSFO for a compound made of two different O isotopes, either $^{16}O$ or $^{18}O$. The chemical potential difference is shown by the green curve and right y-axis in meV/f.u. The vertical dashed line indicates the temperature at which experiments have been performed.

## Oxygen mobility in annealed metal-perovskite

Knowing the mobility and reactivity of oxygen in the pure perovskite system, the annealed metal-perovskite specimens were analyzed to determine how the metallic overlayers influence transport mechanisms. Elemental analyses of the LSFO and LSAT show the phases effectively remain the same as that in the annealed-perovskite case (Figure 2g, h; Table S1). LSAT has not undergone any chemical change or isotopic exchange, again demonstrating it is effectively inert. The bulk LSFO elemental composition remains similar to the annealed-perovskite (and as-grown) specimens, and there is no net oxygen concentration change during reaction (42.3 ± 0.0(3) at.% O). Oxygen isotopic analysis also shows near complete OE in LSFO, similar to that in the perovskite-annealed specimens ($f^{18}O$ 0.82 ± 0.07 vs. 0.84 ± 0.01, respectively).

The elemental and oxygen isotopic composition of the Ni and Cr layers overlying LSFO have changed significantly following annealing, indicative of metal oxidation and phase transformation (Figure 2g, h; Table S1). That is, Ni has clearly oxidized, as the oxygen concentration in the Ni layer has increased from 13.3 ± 1.5 at.% in the as-grown to 44.0 ± 0.4 at.% after annealing, suggestive of a NiO-like phase. The Cr-layer is also partially oxidized following annealing, primarily at its interfaces with other layers; i.e. Cr-rich oxides, similar to $Cr_2O_3$,[53] formed near the Ni/Cr (~45 at.% Cr, 50 at.% O) and Cr/LSFO interfaces (~40 at.% Cr, ~60 at.% O) (see Figure 2). The Cr bulk experiences less oxidation relative to the interfaces and is largely metallic (e.g. only ~25 at.% O), similar to the as-grown material. It is important to note that the specimen outer surface is not included in the APT reconstruction as this was outside the field of view, although we would expect the exposed Cr surfaces to have also oxidized.

Some elemental mixing between the Ni and Cr regions has also occurred, as a thin mixed Ni-Cr oxide layer is now present between the Cr-rich oxide (near the Ni-Cr interface) and the Cr-bulk. In comparison, no cation intermixing (i.e. mixed oxides) is found at the Cr-LSFO interface. Prior work on the oxidation behavior of a Ni-Cr alloys indicated the initial stage of oxidation involved the formation of an outward growing mixed Ni-Cr oxide due to the rapid movement of cations, like Ni.[54–56] Later stages indicate a Cr-rich inward growing oxide layer formed via migration of O from the environment moving towards the metal interfaces and along grain boundaries. This suggests that, at the Ni-Cr interface, a Ni-Cr mixed oxide is initially formed (likely from oxygen originally present in the metal layers; Table S1) followed by a Cr-rich inner oxide.

In addition to cation transport phenomena inferred from elemental analyses, isotopic analyses suggest oxidation occurs by different pathways for each metallic phase (Ni/NiO and Cr/Cr$_2$O$_3$). That is, $f^{18}O$ in the Ni layer has increased from $0.05 \pm 0.02$ in the as-grown specimen to $0.78 \pm 0.07$ after annealing (Figure 2c, i), indicating that oxidation occurs from gaseous $^{18}$O from the environment. Measurements and visualization of the oxygen isotopic distribution also show $^{18}$O homogeneously distributed over the Ni phase (Figure 2i, Figure 4).

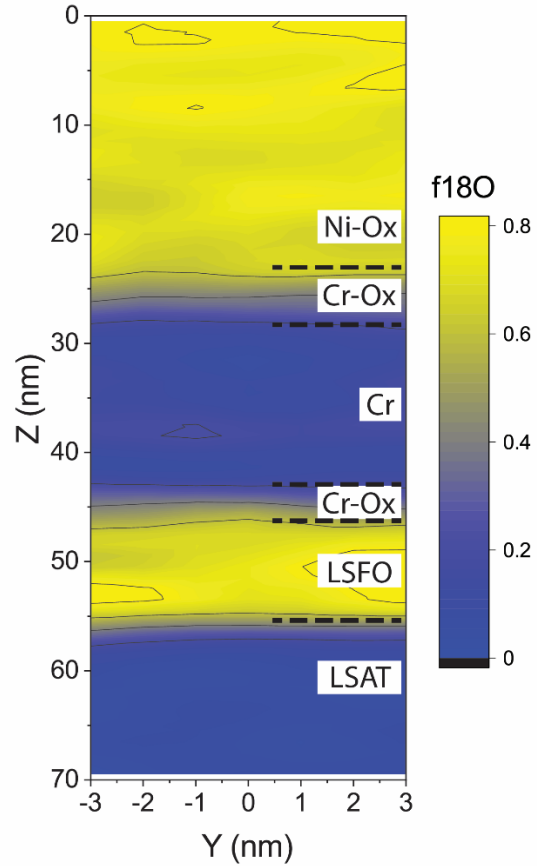

Figure 4: 2D heat map showing the oxygen isotopic composition across the annealed metal-perovskite specimen.

This is not the case for Cr oxidation. That is, the oxygen isotopic content in the bulk is largely $^{16}$O ($f^{18}O =$ $0.15 \pm 0.10$) (Figure 2i, Figure 4). More so, the Cr-oxides formed at the interfaces display a gradient in $^{18}$O, where $f^{18}O$ is highest near the LSFO phases and decreases approaching the Cr bulk (from $0.70 \pm 0.05$ to $0.10 \pm 0.10$). Similar observations are made for the Cr-oxide at the Ni-Cr interface ($f^{18}O$ decreases from $0.38 \pm 0.08$ at the interface to $0.08 \pm 0.04$ towards the Cr-bulk). The $^{18}$O distribution across the interfaces are also noticeably diffuse. Thus, in contrast to Ni oxidation, these observations suggest that Cr within the central volume of the APT specimen, within our field of view, is not simply oxidized by gaseous oxygen from the environment.

Several parameters can contribute to the ability to form a protective, Cr-rich oxide film including thermodynamics of oxide formation, kinetics (i.e., ion mobility), and environmental parameters (oxygen partial pressure, presence of water vapor, temperature, etc.). At the temperature used in this study (400°C), $Cr_2O_3$ is thermodynamically more stable (lower free energy of formation) in comparison to other transition metal oxides such as NiO.[57] Hence, we hypothesize that oxygen anions are mobile and are consumed in the formation of a Cr oxide at the Ni/Cr and Cr/LSFO interface, whereas NiO is formed due to exposure in the *in situ* reactor chamber, where oxygen partial pressure is higher and there is sufficient driving force for Ni oxidation.[58,59] Within the field of view, through the center of the APT specimen, we expect that a Cr oxide is formed from the reaction of Cr with nearby oxygen anions from the neighboring phases (e.g., from migration of oxygen anions from adjacent LSFO and the Ni(-oxide) layer), whereas Ni oxide is formed after exposure to elevated temperature $^{18}O_2$ gas. This is consistent with experimental measurements for the $f^{18}O$ gradient at the Cr-LSFO interface of the annealed specimen. We would expect that outside the field of view, Cr along the surface of the APT specimen oxidizes by the gaseous environment. The passivating Cr-oxide film could prevent $^{18}O$ ingress from the sides of the APT specimen, whereas the Ni (NiO) and LSFO are relatively O permeable. This would make the NiO and LSFO the predominant O sources for Cr oxidation within our field of view.

## Conclusions

Our combined elemental and isotopic observations across the as-grown, annealed perovskite, and annealed metal-perovskite specimens reveal diverse oxygen transport pathways across the system, controlled by the unique reactivity and relationships between phases. This includes (1) exchange between oxygen in LSFO lattice and the gaseous environment, (2) Ni oxidation by gaseous $^{18}O_2$ from the environment, and (3) Cr oxidation by pulling oxygen from the neighboring NiO and LSFO phases (Figure 5).

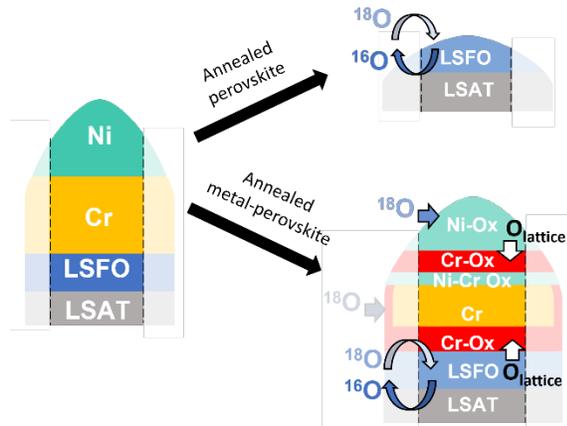

*Figure 5: Conceptual schematic of the diverse oxygen transport pathways and reactions, highlighting OE between lattice oxygen in LSFO and $^{18}O_{2(g)}$ coupled with oxidation of the metallic layers. Vertical dashed lines denote the conceptual field of view captured by APT; we hypothesize that the oxidation and passivation of the Cr could have also occurred over the exposed tip surfaces outside the field of view (i.e., more translucent regions at the shanks).*

Extensive OE in LSFO is consistent with vacancy-mediated mechanisms, where transport and exchange occurs by discrete hops of oxygen ions to neighboring vacant sites and, in this case, leads to the near complete replacement of lattice $^{16}O$ with $^{18}O$ from the environment. The consistent O concentrations in the LSFO for the perovskite-annealed and metal-perovskite annealed specimens relative to the as-grown specimen suggest that steady-state composition is reached, as there is no net addition or removal of oxygen in the LSFO layer (Figure 2). The theoretical simulations performed provide a rationale for OE reaction mechanisms and highlight a complex interplay of different factors. The thermal instability of vacancies, further enhanced by high $O_2$ partial pressure, provides a thermodynamic driver for oxygen atoms from the environment to occupy them. However, the composition (i.e., Sr doping) of the material forces the system to maintain a steady-state vacancy concentration throughout the reaction. While the concentration of O vacancy is dictated by the degree of Sr doping, we propose that defect concentration is the main driver for lattice diffusion and that the increase of the migration energy barrier with Sr doping has a small impact on diffusion and exchange. In addition to thermodynamic of defects and diffusion mechanisms, simulations showed that the isotopic substitution of $^{16}O$ in the LSFO lattice by $^{18}O$ lowers the chemical potential of the material, thereby favoring replacement.

Measurements and visualization of the oxygen isotope distribution at high-spatial and chemical resolution reveal different oxidation mechanisms and pathways for Ni versus Cr. While Cr-species will affect electrode performance by blocking active sites on the surface,[19] we show here that Cr can also potentially scavenge O from neighboring oxides. It also appears the reaction at least partially passivates, as the Cr metal layer did not fully oxidize, although this will depend on the cell's environmental conditions (e.g. temperature, oxygen pressure, etc.).

More broadly, this study demonstrates how the unique coupling of isotopic tracers with APT can provide insight into local, nanoscale reactions as well as potential reaction sequences. While this study was designed to investigate model heterostructures, this approach could be adapted to investigate engineering conditions and materials of interests to SOFCs and beyond. For instance, this approach can be applied to probe oxygen diffusion and exchange systematically, such as across LSFO heterostructures with varying dopants and doping concentrations. This can provide much-needed mechanistic insight into oxygen reactions with perovskites, which are of great significance as electrocatalysts with applications to oxygen reduction reactions. Our approach presents new opportunities to resolve oxygen reaction pathways across complex oxide heterostructures, which can inform fields from catalysis to corrosion science.

## Acknowledgments

This research was supported by the Chemical Dynamics Initiative/Investment, under the Laboratory Directed Research and Development (LDRD) Program at Pacific Northwest National Laboratory (PNNL). DKS acknowledges support from the US Department of Energy (DOE) Office of Science, Basic Energy Sciences, Materials Science and Engineering Division for supporting APT data analysis and interpretation. PNNL is a multi-program national laboratory operated for the U.S. Department of Energy (DOE) by Battelle Memorial Institute under Contract No. DE-AC05-76RL01830. The growth of thin film samples was supported by DOE Office of Science, Basic Energy Sciences under award #10122. Experimental sample preparation and APT analysis was performed at the Environmental Molecular Sciences Laboratory (EMSL), a national scientific user facility sponsored by the Department of Energy's Office of Biological and Environmental Research and located at PNNL. STEM data was collected in the Radiological Microscopy Suite (RMS), located in the Radiochemical Processing Laboratory (RPL) at PNNL. We would also like to thank Dr. Blas Uberuaga (Los Alamos National Laboratory) for helpful discussions on the simulations.

## Data Availability Statement

All relevant data are presented in the main text or supplementary information. STEM, APT, and DFT data can be made available upon request by contacting the authors.

## Author Contributions

S.D.T., K.H.Y., and S.R.S. conceived and developed the project plan. L.W. and Y.D. prepared the thin film samples. S.D.T., K.H.Y., E.J.K., S.N., and D.K.S. conducted APT analysis. B.M. conducted STEM sample preparation and imaging. M.S. conducted theory calculations. All authors contributed to the writing and editing of the manuscript.

# Competing Interests Statement

The authors declare no competing interests.

ToC graphic

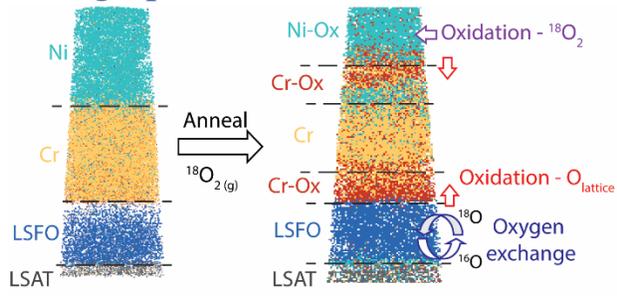

# Supplementary Information for "Resolving diverse oxygen transport pathways across Sr-doped lanthanum ferrite and metal-perovskite heterostructures"

S.D. Taylor,[1,+,*] K.H. Yano,[2,+] M. Sassi,[1] B.E. Matthews,[2] E.J. Kautz,[2] S.V. Lambeets,[1] S. Neumann,[2] D.K. Schreiber,[2] L. Wang,[1] Y. Du,[1] S.R. Spurgeon[2,3,*]

1. Physical and Computational Sciences Directorate, Pacific Northwest National Laboratory, Richland, WA 99352 USA

2. Energy and Environment Directorate, Pacific Northwest National Laboratory, Richland, WA 99352 USA

3. Department of Physics, University of Washington, Seattle, WA 98195 USA

[+]First coauthors.

*Corresponding authors: sandra.taylor@pnnl.gov; steven.spurgeon@pnnl.gov

Includes:

- Figure S1: APT specimen preparation
- Figure S2: Ion assignment in mass spectra
- Table S1: Elemental compositions of the tips measured
- Oxygen isotopic analyses and $f^{18}O$ quantification
    - Figure S3: Analyses of key O ionic and isotopic species
    - Figure S4: $f^{18}O$ quantification
- Figure S5: Analysis of remaining tips

## APT specimen preparation

While samples were prepared in a typical FIB-SEM fashion, additional SEM images are provided here as context for the Ni/Cr-metal capping scheme. SEM images (Figure S1a-b) before sharpening show Pt cap as well as the metal cap, the LSFO, and the LSAT substrate. After sharpening (Figure S1c) the Ni/Cr protective cap remains. For the Annealed specimens, the tip was run through the Ni and Cr layers in the LEAP 4000XHR before annealing. In these cases, the tip would start with the LSFO and have no protective cap remaining.

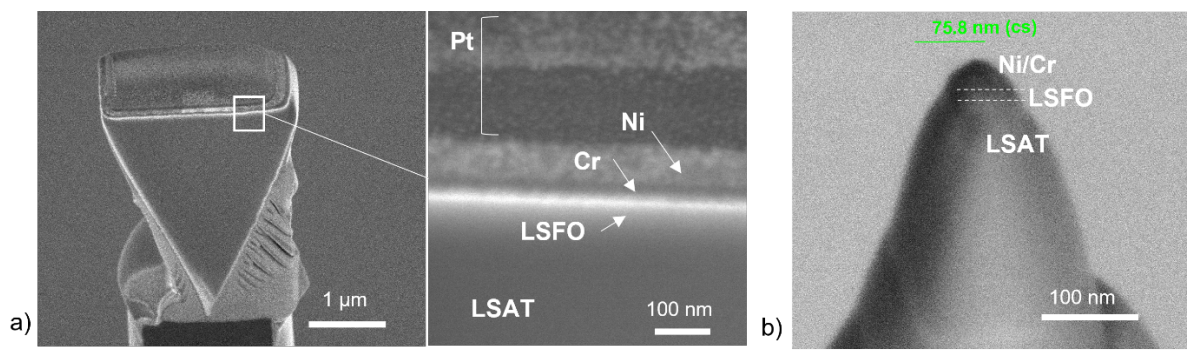

*Figure S1: FIB-SEM preparation of APT tip from the Ni-Cr-LSFO-LSAT stack. Sample preparation is typical for an APT specimen; however, these SEM images are provided for additional context of capping scheme (a) before and (b) after tip sharpening.*

## Elemental analyses

Compositional analyses were enabled through assignment of ionic species to the peaks in the mass spectra consistent with the known elemental and/or isotopic compositions (see description of elemental analysis in SI, Figure S2 and Table S1). Given the multi-elemental nature of this system and resulting complicated mass spectra, the ionic assignment was better enabled by analyzing bulk sections and then deducing major ions appropriate to each phase. In turn, the compositional profiles were calculated by breaking the tip into manageable sections and then stitching them together (Figure S2).

Elemental concentrations were determined by decomposing all ionic species. The error in the concentrations was estimated by standard counting statistics and is represented by the standard deviation σ (Eqn. 1):

$$\sigma = \sqrt{\frac{C_i(100 - C_i)}{N}} \qquad \text{Eqn. 1}$$

where $C_i$ is the measured atomic concentration of element $i$ and $N$ is the total number of atoms detected. Atomic concentrations of each tip condition where bulk concentrations are averaged over two samples with the standard deviation are provided in Table S1. In some cases, LSAT was not captured during the run.

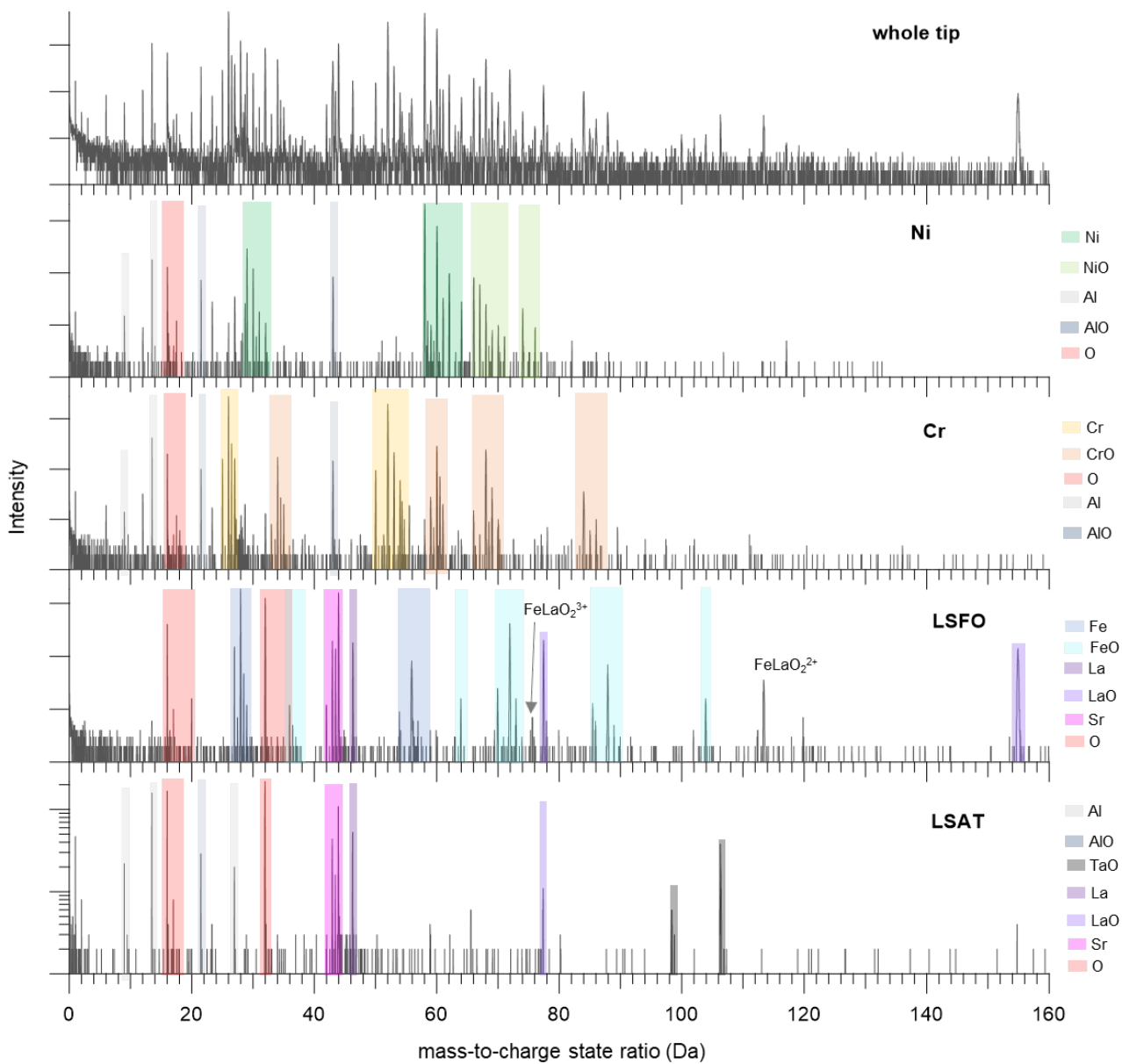

*Fig S2: Mass spectra from as-grown specimen highlighting multi-elemental and isotopic nature of the heterostructure. Ion species were determined through analyses isolating the bulk in each layer. Dominant, general ionic species are labelled. Intensity is log-scale.*

*Table S1: Average elemental and oxygen isotopic compositions of each layer in each set of experiment. Measurements are based off two APT specimens for each experiment, considering the standard deviation (i.e. ± value). LSAT compositions are based off one tip per experiment.*

| At. % | As-Grown | | | | Annealed perovskite | | Annealed metal-perovskite | | | |
|---|---|---|---|---|---|---|---|---|---|---|
| | Ni | Cr | LSFO | LSAT | LSFO | LSAT | Ni | Cr | LSFO | LSAT |
| Ni | 77.29 ± 2.47 | 1.23 ± 0.05 | | | | | 43.43 ± 2.60 | 8.22 ± 1.40 | | |
| Cr | 2.56 ± 0.02 | 70.58 ± 1.53 | | | | | 7.87 ± 0.78 | 50.24 ± 1.43 | | |
| O | 13.33 ± 1.46 | 21.59 ± 1.23 | 43.26 ± 0.91 | 58.65 | 43.79 ± 0.26 | 57.13 | 44.01 ± 0.35 | 34.09 ± 3.44 | 42.31 ± 0.03 | 58.1 |
| La | | | 10.71 ± 0.50 | 7.61 | 10.89 ± 0.02 | 7.15 | | | 11.30 ± 0.22 | 7.2 |
| Fe | | | 27.20 ± 0.11 | | 27.24 ± 0.12 | | | | 27.85 ± 0.02 | |
| Sr | | | 18.49 ± 0.42 | 14.13 | 17.76 ± 0.01 | 13.63 | | | 18.25 ± 0.24 | 13.02 |
| Ta | | | | 8.45 | | 8.04 | | | | 7.47 |
| Al | 6.82 ± 1.46 | 6.59 ± 0.25 | | 10.02 | | 13.19 | 4.70 ± 2.18 | 7.45 ± 0.61 | | 13.13 |
| $f^{18}O$ | 0.05 ± 0.02 | 0.04 ± 0.00(1) | 0.01 ± 0.00(4) | 0.01 | 0.84 ± 0.01 | 0.04 | 0.78 ± 0.07 | 0.19 ± 0.07 | 0.82 ± 0.07 | 0.06 |

## $^{18}O$ analyses and f$^{18}$O quantification

The oxygen isotopic composition was analyzed across the specimen to follow oxygen transport in the solid between the different sources, i.e., $^{18}O_{2\,(g)}$ and the $^{16}O$-solid phases, using the $O^{1+}$ ionic species at 16-18Da. While it was typically a minor oxide-species in all the layers (i.e., ~7 ion% in NiO, ~3 ion% in Cr, ~2 ion% in LSFO, ~9 ion% LSAT), it was consistently observed across the heterostructure – thereby enabling measurements across the interfaces and the different layers. It also led to minimal isobaric and polyatomic interferences, as shown by the f$^{18}$O profile of the as-grown specimen which reproduces concentrations effectively at NA (Figure 2c).

We also assessed the oxygen isotopic composition of all the other major oxide ionic species present in the system for comparison (Fig. S4). We find that most of the major ionic species are isolated to a single phase within the heterostructure (e.g., FeO$^{1+}$ at 70-74 Da is only within LSFO) and thus would not be able to reproduce measurements at interfaces/across the different phases as needed for our study. However, their isotopic compositions are consistent with that of $O^{1+}$ and thus support observations for oxygen exchange or oxidation occurring; for instance, in the LSFO layer, the LaO$^{1+}$ species at 155-157 Da is $^{16}$O-dominant in the as-grown state vs. $^{18}$O-dominant in the annealed (metal-)perovskite specimens, similar to that observed using $O^{1+}$. In some cases, significant interferences were present and complicate isotopic analysis; for instance, $f^{18}O$ analyses using the NiO$^{1+}$ species in the Ni layer (74 – 78 Da) would suffer due unavoidable overlaps between $^{58}$Ni$^{18}$O$^{1+}$ with $^{60}$Ni$^{16}$O$^{1+}$ at 76Da. In turn, these analyses further confirm that oxygen isotopic compositions across the heterostructure are best achieved using the $O^{1+}$ specie.

The oxygen isotopic fraction was calculated from $^{16}$O$^{1+}$ and $^{18}$O$^{1+}$ ion counts (at 16 and 18 Da, respectively) (Eqn. 2). The error or uncertainty in the measured counts was calculated (Eqn. 3), and the uncertainty in the ratios was propagated (Eqn. 4).[60] A bin size of 1 nm was used for the concentration profiles as this led to reasonable measurements in uncertainty while being able to clearly resolve oxygen compositions across the interfaces.

$$f^{18}O = \frac{N_{18O}}{N_{16O} + N_{18O}} = \frac{N_{18O}}{N_O} \qquad \text{Eqn. 2}$$

$$\sigma_x = \sqrt{N_x} \qquad \text{Eqn. 3}$$

$$\sigma_{f18O} = \sqrt{\left(\frac{\sigma_{18O}}{N_{18O}}\right)^2 + \left(\frac{\sigma_o}{N_O}\right)^2} \cdot f^{18}O \qquad \text{Eqn. 4}$$

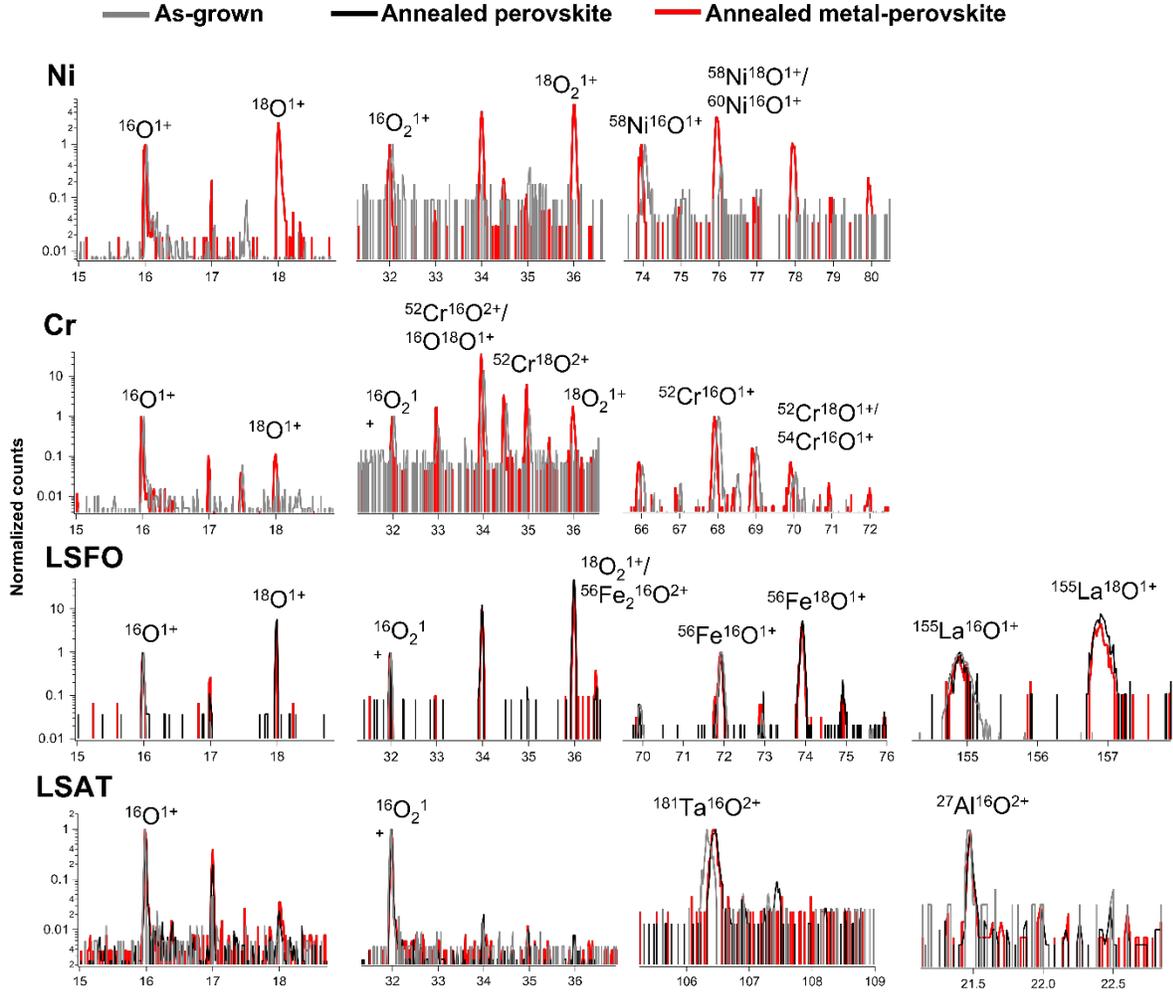

*Figure S3: Isotopic analyses of key O ionic and isotopic species in a) Ni, b) Cr, (c) LSFO, and (d) LSAT phases. Peaks are labelled with their relevant ionic species that are present. Each range is normalized to the dominant $^{16}O$ isotope, to effectively measure changes in $^{18}O$ concentrations from the as-grown to perovskite and metal-perovskite annealed specimens. Potential PMI are identified in some cases, as denoted by multiple ionic species associated with a single peak.*

## Analysis of Remaining Tips

Two tips of each condition were run and the analysis for the second set is provided in Figure S5. Like the data provided in the text (Figure 2), panels (a-c) are of the as-grown, (d-f) the annealed perovskite, and (g-I) the annealed metal-perovskite samples. Atom map reconstructions, elemental concentration profiles, and f18O profiles are all included. These samples are consistent with those presented in the text.

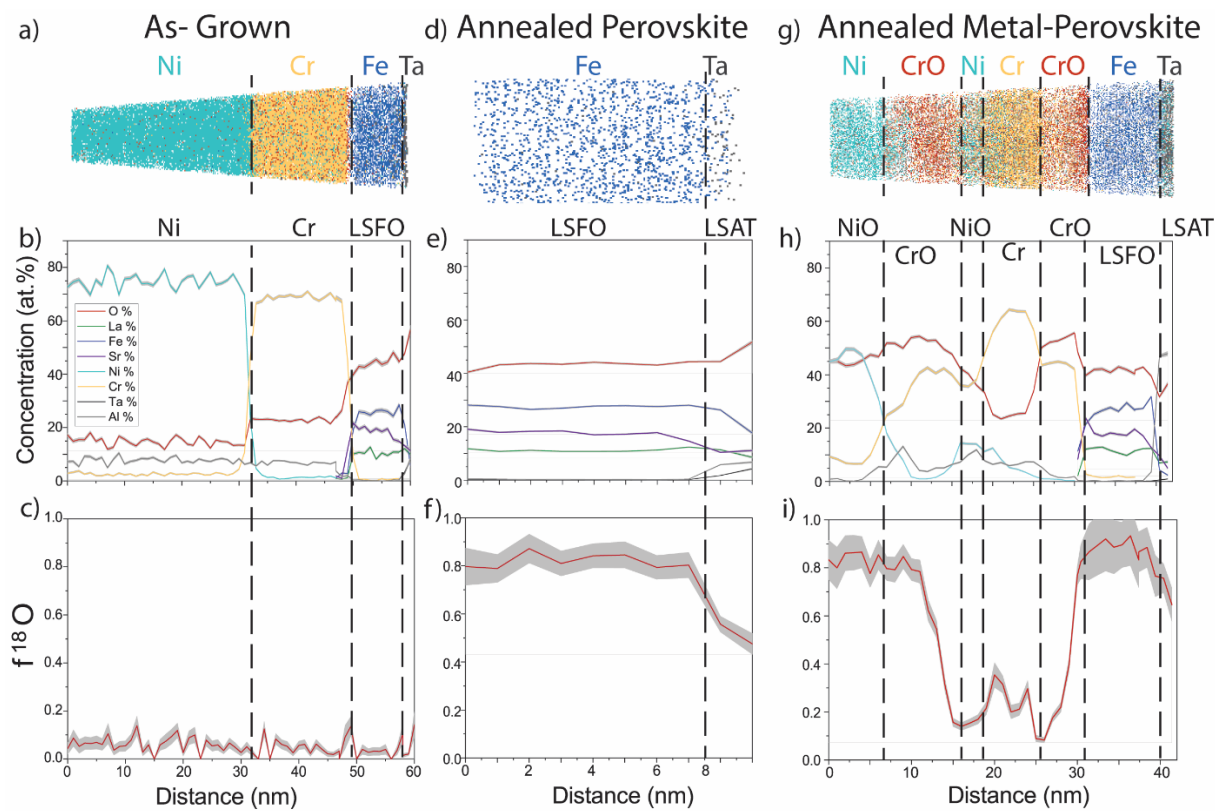

Figure S5. Analyses from additional APT tips with APT reconstructions, elemental profiles, and $f^{18}O$ for the as-grown vs. annealed perovskite vs. annealed metal-perovskite specimens. Shading is used to highlight the uncertainty associated with the counting statistics.